\documentclass[twocolumn,english,aps,pra,nofootinbib,superscriptaddress,showpacs,showkeys,longbibliography]{revtex4-1}
\usepackage[T1]{fontenc}
\usepackage[latin9]{inputenc}
\usepackage{color}
\usepackage{babel}
\usepackage{bbold}
\usepackage{amssymb}
\usepackage{amsmath}
\usepackage{mathtools}
\usepackage{siunitx}
\usepackage{verbatim}
\usepackage{graphicx}
\usepackage{subfigure}
\graphicspath{ {images/} }
\usepackage{bbm, dsfont}
\usepackage{braket}
\usepackage[colorlinks=true,
    linkcolor=red, 
    citecolor=blue, 
    urlcolor =blue]{hyperref}
\makeatletter
\usepackage{amsfonts}
\usepackage{babel}
\usepackage{braket}
\newcommand\minus{
  \setbox0=\hbox{-}
  \vcenter{
    \hrule width\wd0 height \the\fontdimen8\textfont3
  }%
}

\makeatother

\begin{document}

\title{Quantum phase transition with inhomogeneous driving in the Lechner-Hauke-Zoller model}
\author{Andreas Hartmann}
\affiliation{Institute for Theoretical Physics, University of Innsbruck, A-6020 Innsbruck, Austria}
\email{andreas.hartmann@uibk.ac.at}
\author{Wolfgang Lechner}
\affiliation{Institute for Theoretical Physics, University of Innsbruck, A-6020 Innsbruck, Austria}
\email{wolfgang.lechner@uibk.ac.at}
\begin{abstract}
We study the zero-temperature phase diagram of the Lechner-Hauke-Zoller model. An analytic expression for the free-energy and critical coefficients for finite-size systems and in the thermodynamic limit are derived and numerically verified. With the aim to improve standard quantum annealing, we introduce an inhomogeneously driven transverse field with an additional time-dependent parameter that allows one to evade the first-order quantum phase transition and, thus, improve the efficiency of the ground-state preparation considerably.
\end{abstract}
\maketitle

\section{Introduction}
With the recent experimental progress, intermediate scale quantum computers are now available in the laboratory~\citep{Preskill2018quantumcomputingin, Lamata2018, Blatt-Experiment, Neven2016, PhysRevX.6.031007, Lukin, Schoelkopf, PhysRevLett.120.113602, RevModPhys.86.153, GoogleExp, Koch, Walther, Bloch, RevModPhys.73.565}. Although these experiments are not yet ready for scalable quantum computing with error correction, these highly developed platforms are suitable for next generation quantum simulations with full control over the individual degrees of freedom. This has motivated the concept of computation by quantum simulation, i.e. to use the simulation toolbox to solve computational problems. In particular adiabatic quantum computing \citep{RevModPhys.90.015002} (also known as quantum annealing) as a metaheuristic to solve combinatorial optimization problems has been studied extensively \citep{Annealing1, Annealing2, ExperimentAnnealing, Farhi1, AdiabaticQuantumAnnealing, Annealing4, Annealing5, JohnsonQA, Annealing3, DWave, AnnealingTroyer, Rydberg}. Despite the sizable theoretical and experimental efforts, the path towards demonstrating any quantum advantage in adiabatic quantum computing is still elusive. 

From the viewpoint of statistical mechanics, solving optimization problems with adiabatic quantum computing can be understood as driving a random transverse Ising model through a zero-temperature quantum phase transition (QPT) in a one-dimensional phase diagram. In the thermodynamic limit, the QPT is associated with a minimal energy gap that closes exponentially for first-order phase transitions and polynomially for second-order transitions. The limiting factor for the efficiency of quantum annealing is the scaling of the inverse of the squared minimal energy gap between ground and first excited state \citep{XY-EnergyGap, Lidar_2018_FiniteTemperature}. For a random all-to-all model one would expect a second-order phase transition; however, it was recently shown that additional first-order transitions appear in the regime of small transverse fields \cite{knysh2016zero}. 

The Lechner-Hauke-Zoller (LHZ) mapping~\citep{LHZ} is an alternative to the Ising spin glass model. LHZ consists of four-body interactions which are problem independent and random local fields. The associated quantum phase transition is thus fundamentally different compared to the all-to-all spin glass. However, LHZ is similar to an ordered p-spin model with a random local field and one expects a first-order phase transition in the limit where the four-body interactions are dominant. Recently, Susa and co-workers~\citep{Susa2018, PhysRevA.98.042326} (see also Refs.~\cite{Dziarmaga_2010, Adame2018, Rams2016}) showed for $p$-spin models, that inhomogeneneous driving of the transverse field can circumvent the first-order phase transition and improve the ground-state preparation.

In this paper, we derive the expression for the free-energy of the LHZ model and the critical coefficients of the associated quantum phase transition for the thermodynamic limit and estimate finite-size effects. We apply an inhomogeneous driving protocol for the transverse field which has been recently introduced for $p$-spin models \citep{Susa2018, PhysRevA.98.042326} to alter the phase diagram. The inhomogeneous driving introduces an additional dimension in the phase diagram and in this two-dimensional parameter space we are able to evade a particular first-order phase transition in the adiabatic protocol. We numerically demonstrate the implementation of the inhomogeneously driven transverse fields in LHZ and find an enhanced final ground-state fidelity and an enlarged minimal energy gap compared to standard quantum annealing. 

\section{free-energy of the LHZ model}
\subsection{Four-body transverse Ising model}
Transverse Ising models are a cornerstone of modern statistical mechanics and their quantum phase transitions have been studied extensively (for a review, see, e.g., Ref. \cite{stinchcombe1973ising}). The quantum phase transition of random transverse Ising spin models (so-called spin glasses) has recently regained considerable interest with the emergence of quantum annealing as a possible application. Quantum annealing is a metaheuristic that aims at solving combinatorial optimization problems which are encoded in Ising spin glasses \citep{Annealing1, Annealing2}. In this scheme, finding the minimum energy of the spin glass is equivalent to determining the solution of the optimization problem \citep{LucasAnnealing}. In typical examples of encoding optimization problems in the form of Ising models $\mathcal{H}_P = \sum_{i<j} J_{ij} \sigma_{i}^z\sigma_{j}^z$, the interaction matrix $J_{ij}$ has infinite range and is random. 

In quantum annealing, the ground-state of $H_P$ is obtained by adiabatically connecting it to a trivial Hamiltonian, e.g. $\mathcal{H}_I = \sum_k^N \sigma_k^x$. The system is initially prepared in the ground-state of $H_I$ and further sufficiently slowly transferred to the problem Hamiltonian $\mathcal{H}_P$ via the protocol \mbox{$\mathcal{H}(s)=[1-f(s)]\mathcal{H}_I+f(s)\mathcal{H}_P$} where $f(s)$ is a smooth function in the normalized time $s=t/t_f$ with $f(s=0)=0$ and $f(s=1)=1$ and $t_f$ is the running time of the sweep. In switching from $H_I$ to $H_P$, the system undergoes a quantum phase transition which limits its efficiency. For an Ising spin glass, one would expect a second-order phase transition at critical time  $s^*$ with a polynomial closing gap. However, it was recently shown that additional exponentially closing gaps are present for $s>s^*$ \cite{knysh2016zero}. 

An alternative to the spin glass encoding of optimization problems has been recently introduced by LHZ~\citep{LHZ}. In this model, physical qubits describe the relative configuration of two logical spins taking the values $+1$ for parallel (i.e., $\uparrow \uparrow$, $\downarrow \downarrow$) and $-1$ for antiparallel ($\uparrow \downarrow$, $\downarrow \uparrow$) alignment.
The time-dependent Hamiltonian in LHZ reads
\begin{align}
\mathcal{H}_{\textrm{LHZ}}(s)&=\mathcal{H}_I(s) + \mathcal{H}_P(s), \label{eq:eqHamiltonianLHZ} \\
\mathcal{H}_I(s)&=-\sum_{k=1}^{N_p} h_k(s) \sigma_k^x, 
\label{eq:TransverseField} \\
\mathcal{H}_P(s)&=-\sum_{k=1}^{N_p} J_k(s)  \sigma_k^z-\sum_{l=1}^{N_c} C_l(s) \sigma_{l,n}^z\sigma_{l,w}^z\sigma_{l,s}^z\sigma_{l,e}^z,
\label{eq:constraintEnergy}
\end{align}
where $\sigma_k^{x}$ and $\sigma_k^{z}$ are the $x$- and $z$-Pauli matrices for the physical qubit at site $k$ and the strengths of all local fields $h_k$, $J_k$ and constraints $C_l$, respectively, depend on time. Here, \mbox{$\mathcal{H}_I(s)$} is the driver term, and \mbox{$\mathcal{H}_P(s)$} is the encoded problem Hamiltonian to be solved.\\
The strengths of the controllable local magnetic fields $h_k$ and $J_k$ in Equations \eqref{eq:TransverseField} and \eqref{eq:constraintEnergy} are applied to all \mbox{$N_p=N_l(N_l-1)/2$} physical qubits where $N_l$ is the number of logical spins in the original model. The third sum runs over \mbox{$N_c=N_p-N_l+1$} four-body constraints among nearest neighbor qubits on a square lattice and $C_l$ is the strength of a four-body constraint at plaquette $l$. The introduction of these four-body constraints accounts for the increased number of degrees of freedom from $N_l$ logical to $N_p$ physical qubits. This notation excludes \mbox{$N_a=N_l-2$} auxiliary physical qubits in the bottom row of the LHZ architecture to obtain four-body constraints on the whole square lattice. The indices $(l,n)$, $(l,w)$, $(l,s)$, and $(l,e)$ denote the northern, western, southern and eastern physical qubits of the constraint $l$, respectively (more details in Ref. \citep{LHZ}). Given that the constraints are the dominant energy scale, the model is thus similar to the $p$-spin model. 

\subsection{Inhomogeneous transverse field}
The $p$-spin model \citep{Nishimori_NonStoquastic} with a standard  homogeneously driven tranverse field undergoes a first-order quantum phase transition in the zero-temperature phase diagram. Thus, in LHZ--with its similarity to the $p$-spin model for $p=4$--a first-order QPT is also expected during a quantum annealing sweep. As the minimal energy gap of Hamiltonian \eqref{eq:eqHamiltonianLHZ} between the ground-state and the first excited state decreases exponentially with increasing system size $N$ at the critical point (i.e., $\propto e^{-aN}, \, a>0$), the computation time $t_f$ grows exponentially (i.e., $t_f \propto |\langle 1 | d\mathcal{H}/dt | 0 \rangle | /\Delta^2$ with $\Delta$ as the minimal energy gap and $\ket{0}$ and $\ket{1}$ as the instantaneous ground-state and first excited state, respectively) with the system size according to the adiabatic theorem and Landau-Zener's formula.

Spatiotemporal inhomogeneous driving \citep{Mohseni_2018, Susa2018, PhysRevA.98.042326} of the transverse field introduces an additional parameter and, thus, an additional dimension in the phase diagram. This allows one to avoid first-order phase transitions by connecting $H_I$ and $H_P$ via a continuous path around the critical point. This is achieved by switching off the strength of the transverse field inhomogeneously. With the goal to apply this to LHZ in mind, we modify our Hamiltonian \eqref{eq:eqHamiltonianLHZ} as
\begin{align}
\mathcal{H}_{\textrm{LHZ}}(s,r)&=s \mathcal{H}_P(s)-\sum_{k=1}^{N_p}h_k(s,r)\sigma_k^x,
\label{HamiltonianLHZInhomogeneous}
\end{align}
where $h_k(s,r)$ is the strength of the inhomogeneously driven transverse field. In this paper, we choose a protocol for the strength of the transverse field that reads
\begin{align}
h_k(s,r)&=
\begin{dcases}
1 & \textrm{if} \; s < s_{k+1}, \\
N_p(1-s^r)-(N_p-k-1) & \textrm{if} \;s_{k+1} \le s \le s_{k}, \\
0 & \textrm{if} \; s > s_k,
\end{dcases} \nonumber \\
s_k&=[1-(N_p-k)/N_p]^{1/r}, \nonumber \\
s_{k+1}&=[1-(N_p-k-1)/N_p]^{1/r}.
\label{eq:eq42}  
\end{align}
This protocol first switches off the transverse field of the qubits in the first row and the auxiliary qubits in the last row (see Fig.~\ref{fig:figAppendix1} in the Appendix).

The protocol $h_k(s,r)$ is chosen as a continuous piecewise-differentiable function to avoid diverging derivatives of the Hamiltonian \eqref{HamiltonianLHZInhomogeneous}. Here, we have included a new parameter $r$ which enters in an additional time-dependent function $\tau=s^r$ with $0 \le \tau \le 1$. In this spatiotemporal formulation, $s$ and $\tau$ are both controlled as a function of time with $s=\tau=0$ at time $t=0$ and $s=\tau=1$ at time $t=t_f$, the total sweep time.

\begin{figure*}[htb]
\includegraphics[width=0.9\textwidth]{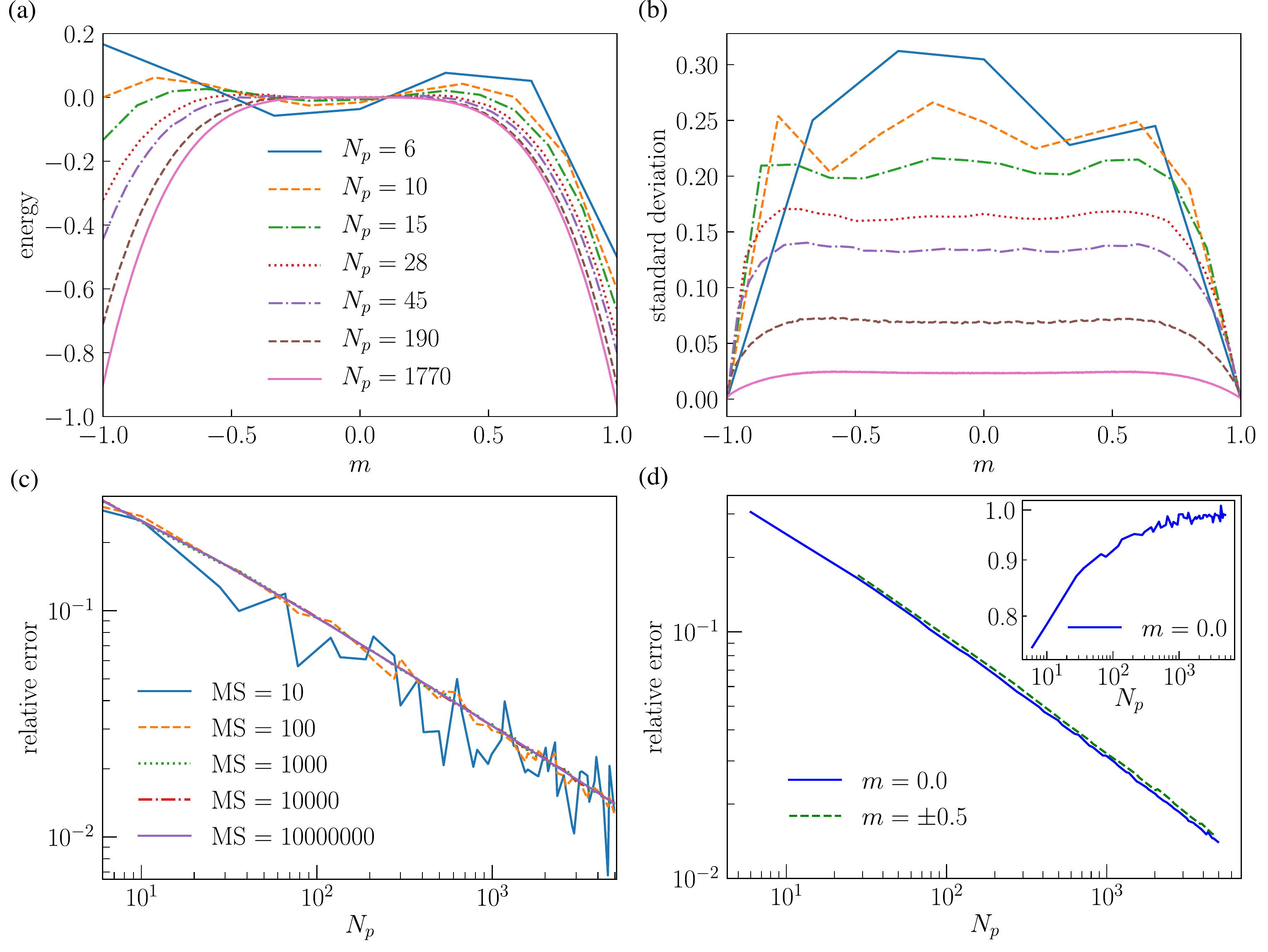}
\caption{\textbf{Energy as a function of the magnetization.}  (a) For small system sizes (blue, upper line), energy from Eq.~\eqref{HamiltonianLHZInhomogeneous} contains an anti-symmetric contribution from the three-body terms and a symmetric contribution form the four-body terms. For large system sizes (pink, lowest line), the symmetric four-body terms dominate the energy. (b) The relative error of the energies decrease with system size and show a maximum at intermediate magnetization. The color code for $N_p$ as in panel (a) from upper descending to the lowest line. (c) The relative error of the energies for magnetization $m=0.0$ as a function of the sampling MS. The relative errors of the energies for magnetizations $m=0.0$ (lower dark gray line) and $m=\pm 0.5$ (upper lighter gray line), respectively, for increasing system sizes $N_p$ are shown in (d). The inset in (d) depicts the relative error multiplied by $\sqrt{N_p}$ for magnetization $m=0.0$ and increasing system sizes. The error arises due to the statistical error of the shuffling. Figures in panel (a), (b) and (d) are performed with MS = 10 000.}
\label{fig:figRelativeEnergyError}
\end{figure*}

\subsection{Inhomogeneously driven LHZ}
In the following, we derive the free-energy of Hamiltonian Eq.~\eqref{HamiltonianLHZInhomogeneous} with an inhomogeneous driving field. This derivation follows the Suzuki-Trotter decomposition used in Refs.~\citep{J_rg_2010, PhysRevE.85.051112}.
The partition function of Hamiltonian Eq.~\eqref{HamiltonianLHZInhomogeneous} reads
\begin{equation}
Z_{\textrm{LHZ}}= \textrm{Tr} \left[e^{-\beta \mathcal{H}_{\textrm{LHZ}}(s, r)}\right].
\label{eq:eq2}
\end{equation}
Using the Suzuki-Trotter decomposition $e^{(A+B)}=\lim_{n \to \infty}\left(e^{A/n}e^{B/n} \right)^n$ with $A,B$ being quantum operators, the partition function reads
\begin{align}
&Z_{\textrm{LHZ}}=\lim_{M \to \infty} Z_{\textrm{LHZ,M}}=\lim_{M \to \infty} \textrm{Tr}\left[e^{-\beta s \mathcal{H}_P/M} e^{-\beta \mathcal{H}_I/M} \right]^M \nonumber \\
&=\lim_{M \to \infty} \sum_{\{\sigma^z \}} \bra{\{ \sigma^z \} } \left\{ \exp\left[\dfrac{s\beta}{M}\sum_{k=1}^{N_p} J_k \sigma_k^z +\dfrac{\beta}{M} \sum_{k=1}^{N_p} h_k \sigma_k^x \right]  \right.\nonumber \\
&\left. \times \exp \left[\dfrac{s \beta}{M}\sum_{l=1}^{\mathrm{N_c}}C_l\, \sigma_{l,n}^z\sigma_{l,e}^z\sigma_{l,s}^z\sigma_{l,w}^z   \right] \right\}^M \ket{\{\sigma^z \} \label{eq:Z}}
\end{align}
where $ \sum_{\{\sigma^z \}} $ refers to the summation over all $2^{N_p}$ possible spin configurations in the $z$-basis and with \mbox{$\{\sigma^z \}=\bigotimes_{k=1}^{N_p} \ket{\sigma_k^z}$}.
We introduce $M$ replicas of the quantum state $\ket{\sigma(\alpha)}$, each labeled $\alpha(=1, \dots ,M)$ such that
$\hat{\mathbb{1}}(\alpha)=\sum_{\{\sigma^z(\alpha)\}} \ket{\{\sigma^z (\alpha)\} }\bra{\{\sigma^z (\alpha)\} } \times \sum_{\{\sigma^x(\alpha)\}} \ket{\{\sigma^x (\alpha)\} }\bra{\{\sigma^x (\alpha)\} }$.
In these replicas, $\alpha$ can be understood as an imaginary time in the dynamic evolution through all these replicas.

Next we derive the expression for the energy of the four-body term in Eq.~\eqref{HamiltonianLHZInhomogeneous}, i.e. $E$ as a function of the magnetization $m$ for a given number of physical qubits $N_p$.\\ LHZ consists of $N_l-2$ three-body plaquettes and \mbox{$(N_l-1)(N_l-2)/2-N_l+2$} four-body plaquettes. Thus, the model resembles a mixture of a $p$-spin model with $p=3$ and $p=4$ and sparse connectivity. Counting the numbers of constraint terms, the energy as a function of magnetization expressed as a function of the number of logical qubits $N_l$ reads as
\begin{equation}
E_{N_l}(m)=-C \left(\dfrac{N_l^2}{2}-\dfrac{5}{2}N_l+3\right)m^4-C (N_l-2)m^3.
\label{eq:energyNl}
\end{equation} 
The same equation expressed as a function of the physical qubits $N_p$ reads as 

\begin{align}
E_{N_p}(m)&=-C (N_p-\sqrt{1+8N_p}+2)m^4 \nonumber \\
&-C(\sqrt{0.25+2N_p}-1.5)m^3.
\label{eq:energy}
\end{align}
Here, we assumed that the constraints are the dominant energy in the system and we neglected the random field terms. Note, that local fields that are drawn randomly form a distribution with mean at $0$ and their contribution to the energy averages out for all magnetizations.  
\\
In order to verify the energy expression \eqref{eq:energy}, we calculate numerically the energy of LHZ averaged over MS samples of each magnetization and fit the result to a function $f(m)=am^4+bm^3$. We repeat this for system sizes between $N_p=6$ up to $N_p=5886$. In order to sample the magnetization, we randomly shuffle the configurations of spins being up or down whereas keeping the total number of spins up constant. For example for the case of the magnetization value $m=0.0$, $N_p/2$ physical qubits are spin-up and $N_p/2$ spin-down. We calculate the energy for randomly shuffled configurations in LHZ with $N_p/2$ qubits being spin-up and $N_p/2$ spin-down, and compute the mean energy and standard deviation of the four-body term. Similarly we proceed for all other possible magnetization values for a chosen number of physical qubits in LHZ.

Figure \ref{fig:figRelativeEnergyError} depicts the energy of the constraints in Eq.~\eqref{eq:constraintEnergy} and its fluctuations using the sampling method described above. The origin of the fluctuations in the energy as a function of magnetization is twofold. One is the result of finite sampling MS, and the other is the result of entropy, i.e., configurations with the same $m$ can have different energies. Let us first consider the error from sampling.

Figure \ref{fig:figRelativeEnergyError}\textcolor{red}{(a)} depicts the energy as a function of the magnetization for various system sizes $N_p$. For small systems, the energy is asymmetric and resembles a cubic function. For large system sizes, the energy approaches a quartic and symmetric function in $m$. The standard deviation of the energies is depicted in Fig.~\ref{fig:figRelativeEnergyError}\textcolor{red}{(b)} for the same system sizes. Figure \ref{fig:figRelativeEnergyError}\textcolor{red}{(c)} depicts the relative errors as a function of the shuffling parameter MS. The standard deviation scales, as expected with the system size as $1/\sqrt{N_p}$. For small and large values of shuffling parameter MS, we plotted the relative error of the energy for magnetization $m=0.0$ for different system sizes $N_p$.  Figure \ref{fig:figRelativeEnergyError}\textcolor{red}{(d)} shows the relative error for magnetizations $m=0.0$ and $m=\pm0.5$ and different system sizes $N_p$. The inset in Fig.~\ref{fig:figRelativeEnergyError}\textcolor{red}{(d)} shows the relative error multiplied by the inverse of the scaling, i.e. $\sqrt{N_p}$, for magnetization $m=0.0$ and increasing system sizes $N_p$. Thus for small systems, the entropic energy fluctuations persist for all system sizes.

\begin{figure}
\centering
\includegraphics[width=0.45\textwidth]{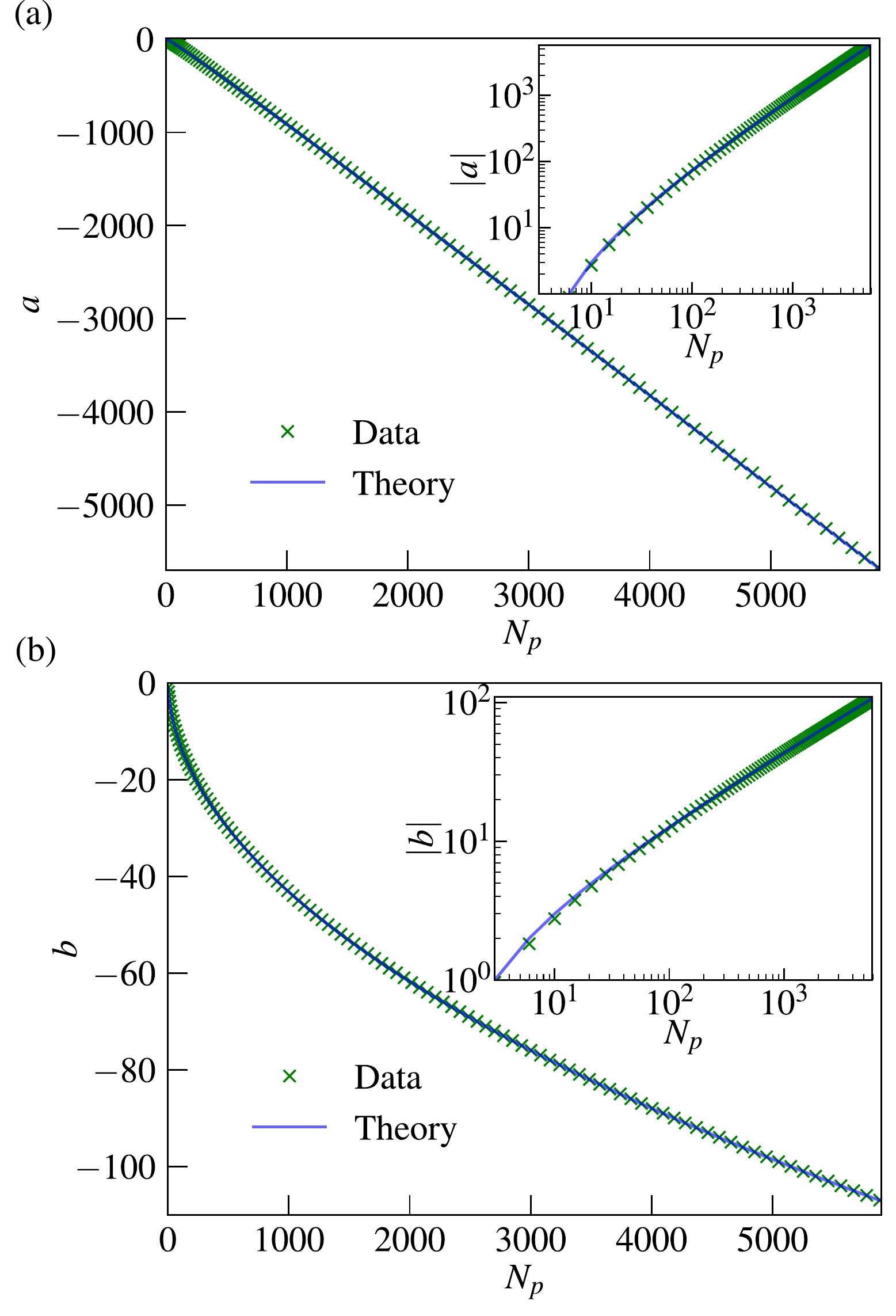}
\caption{\textbf{Energy of the constraints.} The energy in Eq.~\eqref{eq:energy} contains a cubic $m^3$ and a quartic $m^4$ term, both functions of the system size $N_p$. The analytical expression for the cubic term $a$ is shown in panel (a) and for the quartic term $b$ in panel (b) [both as solid blue (dark gray) line]. For comparison, the numerical data are depicted with green (gray) cross symbols. The inset shows the same data on the logarithmic scale of the absolute values. Note that there are no fit parameters used in this figure.}
\label{fig:figFitParameters}
\end{figure}

With the numerical data for the energy, we can now verify the individual terms in the analytical expression of Eq.~\eqref{eq:energyNl}. Figure \ref{fig:figFitParameters} depicts the comparison of Eq.~\eqref{eq:energy} and the numerical data with a fit $f(m)=am^4+bm^3$ with parameters $a$ and $b$. The analytical expressions from Eq.~\eqref{eq:energy} are in excellent agreement with the data for both, the cubic and the quartic terms. 

As expected from the $N_p$ dependence in Eq.~\eqref{eq:energy}, the quartic $m^4$ term dominates in the thermodynamic limit as shown in Fig.~\ref{fig:figAppendixEnergyTerm}\textcolor{red}{(b)} in the Appendix for the case of $N_l=109$ logical and, thus, $N_p=5886$ physical qubits in LHZ.\\

Plugging expression Eq.~\eqref{eq:energy} into Eq.~\eqref{eq:Z}, the decomposed partition function $Z_{\textrm{LHZ,M}}$ reads with the help of the integral definition of the $\delta$ function, i.e., $\delta(N_p m(\alpha)-\sum_{k=1}^{N_p}\sigma_k^z(\alpha))=\int d\tilde{m}(\alpha) \exp[-\tilde{m}(\alpha)(N_p m(\alpha)-\sum_{k=1}^{N_p}\sigma_k^z(\alpha))]$, as
\begin{align}
&Z_{\textrm{LHZ,M}}= \nonumber \prod_{\alpha=1}^M \int d m(\alpha) d \tilde{m}(\alpha) \nonumber \\
&\exp \left\{\sum_{\alpha=1}^M \dfrac{\beta s C}{M} \left[\left(N_p-\sqrt{1+8N_p}+2 \right) m(\alpha)^4 \right.\right. \nonumber \\
&\left.\left.+\left(\sqrt{0.25+2N_p}-1.5\right) m(\alpha)^3 \right]-N_p \tilde{m}(\alpha) m(\alpha) \right\} \times \nonumber \\
&\exp \left\{\sum_{k=1}^{N_p} \ln \mathrm{tr} \prod_{\alpha=1}^M \exp\left[\left(\tilde{m}(\alpha)+\dfrac{\beta s}{M}J_k\right)\hat{\sigma}^z\right] \exp \left[\dfrac{\beta}{M} h_k \hat{\sigma}^x \right] \right\} \nonumber \\
&=\prod_{\alpha=1}^M \int d m(\alpha) d \tilde{m}(\alpha) \exp[-N_p \beta f_{N_p,M}(\{m(\alpha)\})],
\end{align}
where $f_{N_p,M}(\{m(\alpha)\})$ is the free-energy of the system consisting of $N_p$ physical qubits as a function of the magnetization $m$. The parameter $M$ denotes the number of imaginary time slices and $\beta$ is the reciprocal temperature.\\

We are interested in an expression for $\tilde{m}$ that minimizes the integrand of the partition function and thus the free-energy of LHZ. This saddlepoint condition for $\tilde{m}(\alpha)$, i.e. $\partial Z_{\textrm{LHZ,M}}/\partial m=0$ and solving for $\tilde{m}$, reads 
\begin{align}
\tilde{m}(\alpha)&=\dfrac{\beta s C}{M} \left[\left(4-\dfrac{\sqrt{16+128N_p}+8}{N_p}\right)m(\alpha)^3 \right. \nonumber \\
&\left.+\dfrac{\sqrt{2.25+18N_p}-4.5}{N_p} m(\alpha)^2 \right],
\end{align}
and the free-energy, thus, becomes
\begin{align}
& f_{N_p,M}(\{m(\alpha)\})=\dfrac{sC}{M} \sum_{\alpha=1}^M \left(3 + \dfrac{6-\sqrt{9+72N_p}}{N_p}\right) m(\alpha)^4\nonumber \\
&+\left(\dfrac{\sqrt{1+8N_p}-3}{N_p}\right)m(\alpha)^3 \nonumber \\
&-\dfrac{1}{\beta N_p} \ln \mathrm{tr} \prod_{\alpha=1}^M \exp \left\{\dfrac{\beta s}{M}\left[C\left(4-\dfrac{\sqrt{16+128N_p}+8}{N_p}\right) m(\alpha)^3 \right.\right. \nonumber \\
&\left.\left. +C\left(\dfrac{\sqrt{2.25+18N_p}-4.5}{N_p} \right)m(\alpha)^2+J_k \right]\hat{\sigma}^z \right\} \nonumber \\
& \times \exp \left(\dfrac{\beta}{M}h_k \hat{\sigma}^x \right).
\label{eq12:freeenergy}
\end{align}
We now apply the static approximation $m=m(\alpha)$ for all $\alpha$ and take the reverse operation of the Suzuki-Trotter decomposition for $M \to \infty$. We can further rewrite the strength $h_k$ of the inhomogeneously driven transverse field as a continuous function $h(\tau')$ with $\tau' \in [0, 1]$ as denoted in Eq.~\eqref{eq:eq42}
to obtain the integral form of the free-energy in the zero-temperature limit $\beta \to \infty$, i.e., $1/\beta \ln \, 2 \, \cosh \, \beta \to 1$, as
\begin{widetext}
\begin{align}
f(m,s,\tau',C,J,N_p)&=sC\left[\left(3+\dfrac{6-\sqrt{9+72N_p}}{N_p} \right)m^4+ \left(\dfrac{\sqrt{1+8N_p}-3}{N_p} \right) m^3 \right] \nonumber \\
&-\left[\int_0^1 d\tau'\sqrt{s^2\left(C\left(4-\dfrac{\sqrt{16+128N_p}+8}{N_p}\right)m^3+C\left(\dfrac{\sqrt{2.25+18N_p}-4.5}{N_p} \right)m^2+J\right)^2+h(\tau')^2}\right],
\label{eq:eqfreeenergytheory}
\end{align}
\end{widetext}
where the square brackets $[\cdots]$ over the integral denote the average value over the distribution of the strengths $J_k$ of the longitudinal magnetic field denoted as $J$ according to the law of large numbers \mbox{$\lim_{N_p \to \infty} 1/N_p \sum_{k=1}^{N_p}(\cdots)=[(\cdots)]$}. This means  the integral is evaluated for uniformly distributed values of $J_k$. Equation \eqref{eq:eqfreeenergytheory} together with the protocol $h(s,r)$ in Eq.~\eqref{eq:eq42} for the inhomogeneously driven transverse field [i.e., $h(\tau')$ in Eq.~\eqref{eq:eqfreeenergytheory}] describes the free-energy of LHZ for finite sizes with respect to the number $N_p$ of physical qubits (see the Appendix for further information).

\begin{figure}
\centering
\includegraphics[width=0.5\textwidth]{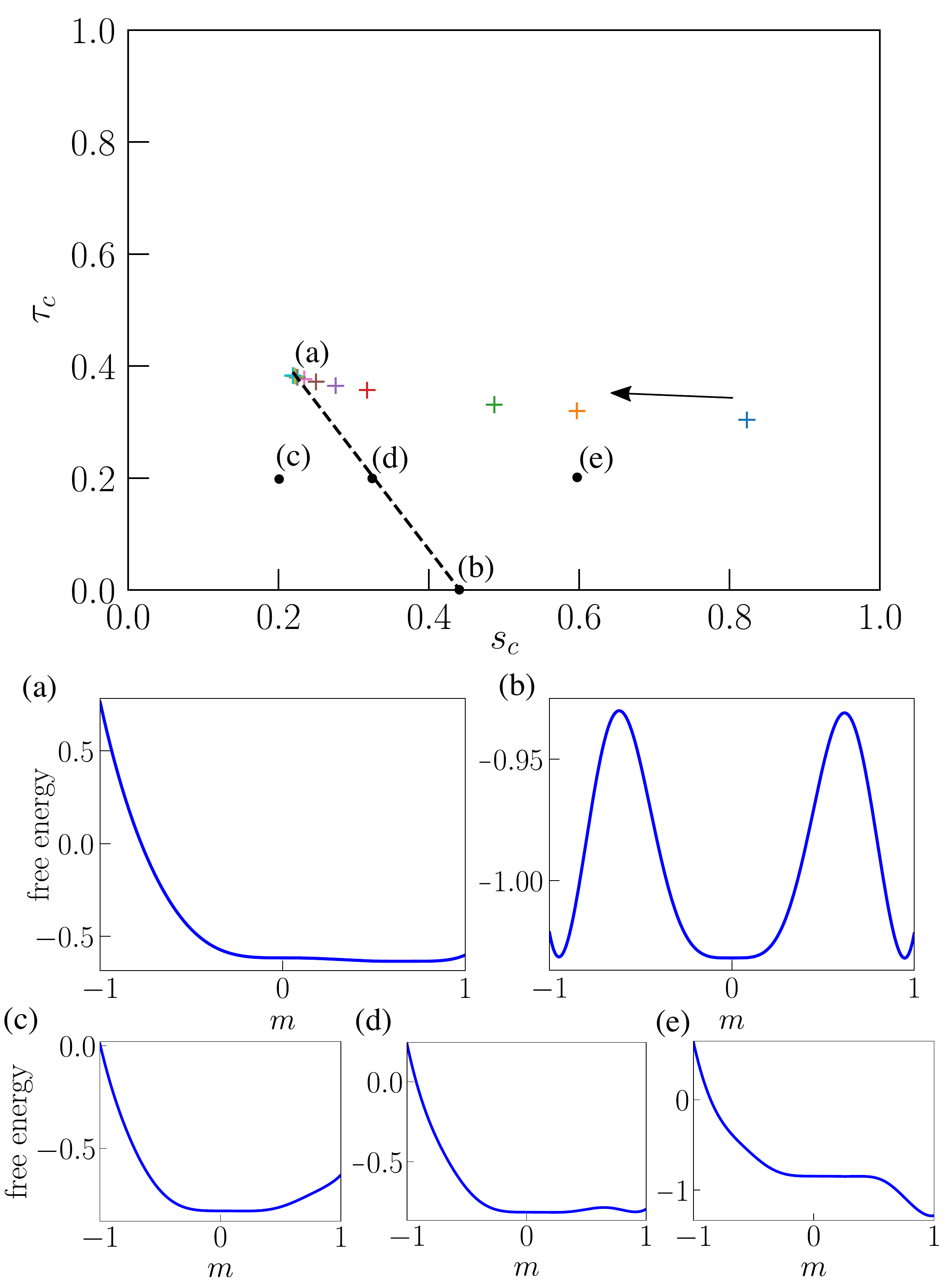}
\caption{\textbf{Critical coefficients.} The evolution of the critical coefficients $\tau_c$ and $s_c$ of Eq.~\eqref{eq:eqfreeenergytheory} for increasing number of physical qubits from $N_p=28$ [blue (dark gray) plus on the right)]to $N_p=499500$ (gray plus at (a)) is shown. In the thermodynamic limit, we reach the critical coefficients of Eq.~\eqref{eq:eqCriticalCoefficients} of the thermodynamic free-energy term Eq.~\eqref{eq:freeEnergy} at (a). The subplots (a)--(e) show the free-energy with respect to the magnetization $m$ for different values of $s$ and $\tau$ of Eq.~\eqref{eq:freeEnergy}.}
\label{fig:figCriticalCoefficients}
\end{figure}

For finite-size systems with $N_p$ physical qubits in LHZ, we use the expression $h(s,\tau)$ of Eq.~\eqref{eq:eq42} for $h(\tau')$ in Eq.~\eqref{eq:eqfreeenergytheory}.
In the thermodynamic limit $N_p \gg 1$, the interval $[s_{k+1},s_k]$ for $s$ of Eq.~\eqref{eq:eq42} becomes infinitesimally small, and thus, the protocol for the strength of the inhomogeneously driven transverse field can be written as
\begin{equation}
h(\tau')=
\begin{dcases}
1 & \textrm{for} \; 0 < \tau' < 1-\tau, \\
0 & \textrm{for} \ 1-\tau < \tau' < 1.
\end{dcases}
\label{eq:eq41}
\end{equation}
For this choice of inhomogeneous transverse field strength $h$, the free-energy of LHZ in the thermodynamic limit can be written as
\begin{align}
f(m,s,\tau,C,J)&=3sCm^4+\left[-\tau s (4Cm^3+J)\right. \nonumber \\
&\left.-(1-\tau)\sqrt{s^2(4 C m^3+J)^2+1}\right],
\label{eq:freeEnergy}
\end{align}
where $C$ and $J$ are the strengths of the constraints and longitudinal magnetic fields, respectively.\\
The critical coefficients $m_c$, $s_c$, and $\tau_c$ of the free-energy term Eq.~\eqref{eq:freeEnergy} are obtained by a Landau-type expansion of the free-energy term with the condition that the first three derivatives vanish \citep{nishimori2011elements}, i.e., we have to solve the system of equations
\begin{subequations}
\begin{align}
&\dfrac{\partial}{\partial m}f(m,s,\tau,C,J)\bigg|_{m=m_c}=0, \\
&\dfrac{\partial^2}{\partial m^2}f(m,s,\tau,C,J)\bigg|_{m=m_c}=0, \\
&\dfrac{\partial^3}{\partial m^3}f(m,s,\tau,C,J)\bigg|_{m=m_c}=0,
\end{align}
\label{eq:eqEqSys}
\end{subequations}
with respect to its critical coefficients $m_c$, $s_c$ and $\tau_c$. Note, that the critical coefficients are a function of the constraint strength $C$ and the distribution of $J$.
The values of the critical coefficients for a uniform distribution of the strength of the longitudinal magnetic field $J$ with values between $-1$ and $1$, and constraint strength $C=2$ are 
\begin{equation}
m_c \approx 0.679\,795, \; s_c  \approx 0.219\,232, \; \tau_c \approx 0.389\,11.
\label{eq:eqCriticalCoefficients}
\end{equation}
We can obtain these critical coefficients with our thermodynamical free-energy term Eq.~\eqref{eq:freeEnergy} [or with finite-size free-energy term Eq.~\eqref{eq:eqfreeenergytheory} by increasing the number of physical qubits $N_p$] and which can be seen in Fig.~\ref{fig:figCriticalCoefficients}. Here, we have plotted the free-energy term of Eq.~\eqref{eq:freeEnergy} with respect to the magnetization $m$ for different points $(s,\tau)$ in the two-dimensional phase diagram. On the first-order transition line [points (a), (b) and (d)], we see a degenerated minimum of the free-energy. At the crossing of the first-order transition line starting from (c) and going to (e), we see that the value for the magnetization that minimizes the free-energy changes discontinuously from the paramagnetic solution $m=m_\mathrm{p}=0.0$ to the ferromagnetic solution $m=m_\mathrm{f}=1$ and which depicts a quantum phase transition of first order.

\section{Numerical Results}
Let us now apply the results to a quantum annealing protocol. For quantum annealing, an important measure of the efficiency is the  ground-state fidelity $F(t_f)=\langle \psi(t_f) | \phi_0(t_f) \rangle$ with $\ket{\psi(t_f)}$ as the state of our system and $\ket{\phi_0(t_f)}$ as the ground-state of our final Hamiltonian at time $t=t_f$. Another important measure is the minimal energy gap $\Delta E_{\textrm{min}}$ of the corresponding energy eigenspectra.

\begin{figure}
\centering
\includegraphics[width=0.5\textwidth]{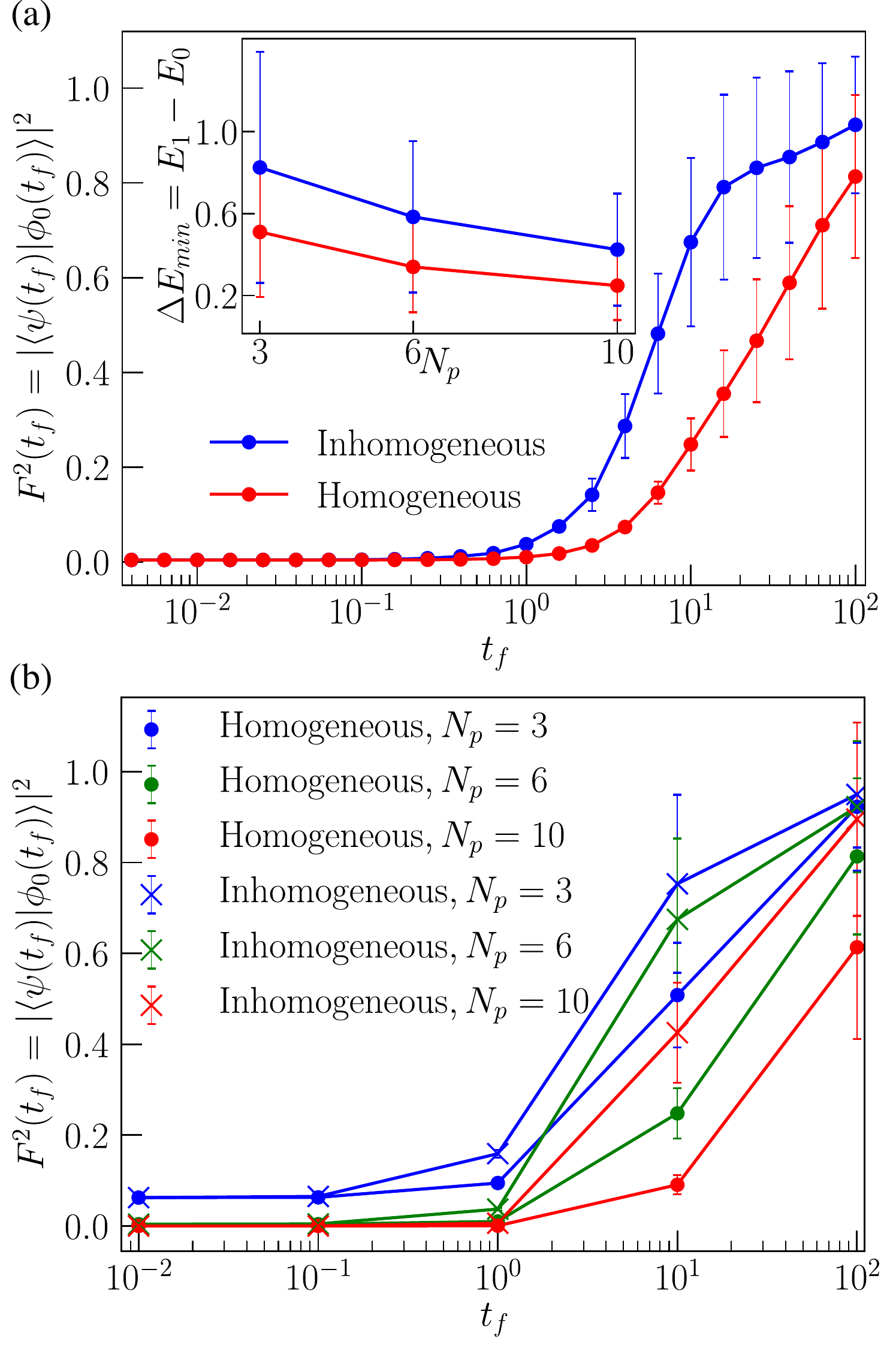}
\caption{\textbf{Final ground-state fidelities and minimal energy gaps for sweeps with homogeneous and inhomogeneous driving.}  (a) shows the statistics of the squared final ground-state fidelities for an ensemble of 100 instances with system size $N_l=4$ logical qubits and, thus, $N_p=6$ physical qubits plus two auxiliary qubits with uniformly distributed interaction strengths $J_k$ and constraint strength $C=2$. (b) shows the comparison of the squared final ground-state fidelities of homogeneous \eqref{eq:eqHamiltonianLHZ} and inhomogeneous Eq.~\eqref{HamiltonianLHZInhomogeneous} drivings with parameter $r=0.5$. The inset in (a) depicts the statistics of the minimal energy gap of these chosen uniformly distributed instances $J_k$ for different system sizes.}
\label{fig:figFidelities}
\end{figure}

Figure \ref{fig:figFidelities} shows the statistics of the squared final ground-state fidelities $F^2(t_f)$ for sweeps with different running times $t_f$ for an ensemble of 100 randomly uniformly distributed instances of interactions $J_k$ for homogeneous \eqref{eq:eqHamiltonianLHZ} and inhomogeneous Eq.~\eqref{HamiltonianLHZInhomogeneous} drivings, respectively. The system size is $N_p=6$ plus two auxiliary physical qubits, and strengths of the constraints are each $C=2$ for all three constraints and $10$ is the value of the strength of the auxiliary local fields in the bottom row of LHZ. The free parameter in the inhomogeneously driven transverse field is $r=0.5$.
One can see that inhomogeneous driving of the transverse field can enhance the performance of traditional quantum annealing considerably. Furthermore, the ratio of the squared final ground-state fidelities of inhomogeneous to homogeneous driving improves with increasing system size as shown in Fig.~\ref{fig:figFidelities}\textcolor{red}{(b)}. 

A free parameter in the protocol is the choice of the control parameter $r$ in the inhomogeneous sweep, i.e., the path we take in the two-dimensional $s-\tau$ diagram. Figure \ref{fig:figComparisonR} in the Appendix shows that the control parameter $r$ can considerably enhance the performance of inhomogeneous driving.\\

The excess energy can also be lowered considerably by inhomogeneous driving of the transverse field. For the same ensemble of 100 randomly chosen instances for system size $N_p=6$ plus two auxiliary physical qubits, we have plotted the excess energy $E=\sum_n E_n-E_0$ (where $E_n$ denotes the energy of the $n$-th lowest eigenstate) for homogeneous \eqref{eq:eqHamiltonianLHZ} and inhomogenous \eqref{HamiltonianLHZInhomogeneous} driving of the transverse field in Fig.~\ref{fig:figExcessEnergies} in the Appendix. 

An interesting question arises whether the particular order at which the transverse fields of the qubits are switched off is relevant for the performance. Statistics of the ground-state fidelities and excess energies of homogeneous \eqref{eq:eqHamiltonianLHZ} and inhomogeneous Hamiltonian \eqref{HamiltonianLHZInhomogeneous} with value $r=0.5$ and descending order are included in Fig.~\ref{fig:figStatisticsDescending} in the Appendix. The ground-state fidelities and excess energies are the same as for the ascending order case. The method is insensitive to the precise order in which the transverse fields are switched off. 

The minimal energy gap is considered one of the main limiting factors in quantum annealing. The inset of Fig.~\ref{fig:figFidelities}\textcolor{red}{a} depicts the statistics of the minimal energy gaps of homogeneous Hamiltonian \eqref{eq:eqHamiltonianLHZ} and inhomogeneous Hamiltonian \eqref{HamiltonianLHZInhomogeneous} with value $r=0.5$ for different system sizes $N_p$ over an ensemble of 100 randomly chosen instances of interaction strength $J_k$.\\
Due to inhomogeneous driving of the transverse field we can considerably enlarge the minimal energy gap for all instances compared to standard quantum annealing. Also, the ratio of the minimal energy gap of inhomogeneous driving to homogeneous driving increases with system size.

\section{Conclusion and Outlook}
In this paper, we have introduced an inhomogeneously driven transverse field of the Hamiltonian in the LHZ lattice gauge model architecture. We find that by using inhomogeneous driving of the transverse field in LHZ the ground-state fidelities are increased considerably compared to standard quantum annealing. The method is insensitive to the order in which the fields are switched off. 

As an important step, we analytically derived an energy expression of the four-body constraint term $\sigma^z \sigma^z \sigma^z \sigma^z$ in LHZ. The term in front of the $m^4$ term stems from the four-body constraints of LHZ whereas the term in front of $m^3$ stems from the three-body constraints in the lower row of LHZ. As the ratio of three-body to four-body constraints converges towards the value 0 for increasing system sizes $N_p$, i.e., the finite-size effect of the three-body constraints vanish, the $m^4$ term dominates as can be seen in Figs.~\ref{fig:figRelativeEnergyError}\textcolor{red}{(a)} and \ref{fig:figAppendixEnergyTerm} in the Appendix. In our derivation, we followed the Suzuki-Trotter decomposition, the saddlepoint approximation as well as the static approximation. We believe that these approximations are valid for LHZ as it was shown recently that it reproduces the exact free-energy term under some valid constraints \citep{okuyama2018exact}.

From our free-energy expression \eqref{eq:eqfreeenergytheory} of LHZ, we calculated the critical coefficients for different system sizes and for the thermodynamic limit, respectively, and where we further computed the line of first-order quantum phase transitions. We note here that, for small system sizes, we can always avoid first-order phase transitions for different values of the control parameter $r$, i.e., taking different paths through the two-dimensional $s-\tau$ diagram. We further note that, for quantum annealing in LHZ for larger system sizes, the strength of the constraints obey a scaling behavior which changes the values of its critical coefficients considerably. We have included an analytical expression of the free-energy expression for a scaling behavior of $C \propto \sqrt{N_p}$ and the calculation of the critical coefficients for different system sizes in the Appendix (see also Fig.~\ref{fig:AppendixCriticalCoefficients}).

Furthermore, we numerically demonstrated an increase in the minimal energy gap and final ground-state fidelity due to avoiding first-order quantum phase transitions by inhomogeneously driving the strength of the transverse field. We note that the ratio of the final ground-state fidelities of inhomogeneous driving Eq.~\eqref{HamiltonianLHZInhomogeneous} to homogeneous driving Eq.~\eqref{eq:eqHamiltonianLHZ} increases with the system size $N_p$ of physical qubits in LHZ which we expect due to the exponential closing of the minimal energy gap with system size. This is an encouraging result which we will further study numerically for larger system sizes by using path integral Monte Carlo methods \citep{Troyer1, Troyer2, Inack2018}  in future work.

We note, that the minimal gap of random Ising models is not directly at the critical point but rather in the spin-glass phase close to the end of the adiabatic protocol \cite{knysh2016zero}. Whether inhomogeneous driving also affects these additional exponential closing gaps is an interesting question for future studies. 

As a future direction, the inhomogeneous driving scheme in the LHZ model may be applied to the counter-diabatic driving of LHZ as described in Ref.~\citep{Hartmann_2019} where quantum phase transitions during the sweep may decrease the efficiency of the approximate counterdiabatic term added to speedup quantum annealing. This may open a new branch of developing fast near-term quantum annealer devices.

\section{Acknowledgments}
We thank T. Hatomura, H. Nishimori, L. Sieberer and K. Ender for valuable discussions. The research was funded by the Austrian Science Fund (FWF) through a START Grant under Project No. Y1067-N27 and the SFB BeyondC Project No.~F7108-N38, the Hauser-Raspe Foundation, and the European Union's Horizon 2020 Eesearch and Innovation Program under Grant Agreement No.~817482 PasQuanS.

\newpage 

\newpage

\appendix*
\numberwithin{equation}{section} 
\numberwithin{figure}{section} 
\section*{Appendix}
\setcounter{equation}{0}
\setcounter{figure}{0}
\renewcommand{\theequation}{A\arabic{equation}}
\renewcommand{\thefigure}{S\arabic{figure}}

\subsection*{Inhomogeneous driving of the transverse field}
\begin{figure}[htb]
\centering
\includegraphics[width=0.5\textwidth]{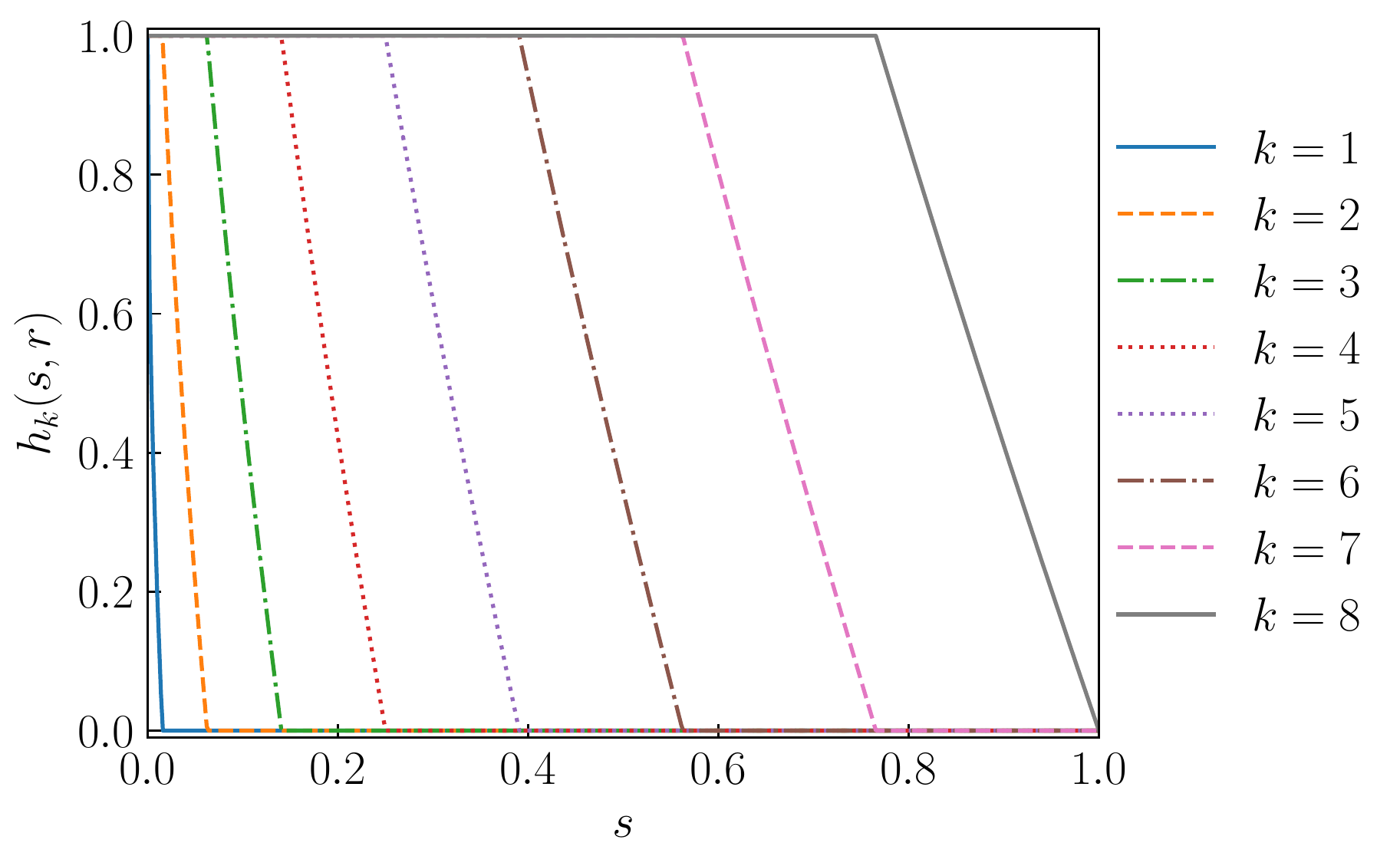}
\caption{\textbf{Continuous function $h_k(s,r)$.} The value of the continuous piecewise function $h_k(s,r)$ of Eq.~\eqref{eq:eq42} in the main text over the normalized time $s=t/t_f$ for $N_p=6$ plus two auxiliary physical qubits and parameter value $r=0.5$ as used for the results in the main text are shown. The blue (dark gray) solid line corresponds to the protocol of the first qubit and similar in ascending order and the gray solid line for the last qubit in the bottom row of LHZ, i.e., $k=8$.}
\label{fig:figAppendix1}
\end{figure}

The inhomogeneous driving of the transverse field $h_k(s,r)$ in the main text is assumed to be a linear function with different slopes and delays. Figure \ref{fig:figAppendix1} shows the function $h_k(s,r)$ of Eq.~\eqref{eq:eq42} in the main text for \mbox{$N_l=4$} logical and, thus, $N_p=6$ plus two auxiliary physical qubits in the bottom row in LHZ with parameter value $r=0.5$ in the inhomogeneous driving scheme. Here, $s=t/t_f$ is the normalized time, and $k=1$ denotes the physical qubit on the lower left in LHZ and $k=8$ denotes the additional physical qubit fixed to a value of 1 on the bottom right in LHZ, i.e., we first switch off the transverse field of the physical qubit in the lower left and at last the additional qubit in the bottom right.

\subsection*{Derivation of the free-energy in LHZ}
The free-energy of LHZ in the thermodynamic limit $N_p \to \infty$ [Eq.~\eqref{eq:freeEnergy} in the main text] is similar to the free-energy expression of the $p$-spin model for $p=4$ \citep{PhysRevA.98.042326, Susa2018}. Compared to the $p$-spin model, the LHZ model contains $m^4$ and $m^3$ terms with a ratio that depends on the system size. In LHZ, the ratios $N_4/N_c$ of the number of four-body constraints to all constraints as well as the ratios $N_3/N_c$ the number of three-body constraints to all constraints read
\begin{align}
F_4&=\dfrac{N_4}{N_c}=1-\dfrac{N_l-2}{\dfrac{N_l^2}{2}-\dfrac{3}{2}N_l+1}, \nonumber \\
F_3&=\dfrac{N_3}{N_c}=\dfrac{N_l-2}{\dfrac{N_l^2}{2}-\dfrac{3}{2}N_l+1}.
\end{align}
These scalings translate to the terms in front of $m^4$ and $m^3$ [i.e., Equations \eqref{eq:energyNl} and \eqref{eq:energy}] compared to the $p$-spin model with $p=4$.
The ratio $F_4$ of four-body constraints to all constraints converges towards 1 and the ratio \mbox{$F_3=1-F_4$} towards 0, meaning that the finite-size effect of three-body constraints are negligible in the thermodynamic limit and vice versa the term of the four-body constraints dominates.
\begin{figure}
\centering
\includegraphics[width=0.45\textwidth]{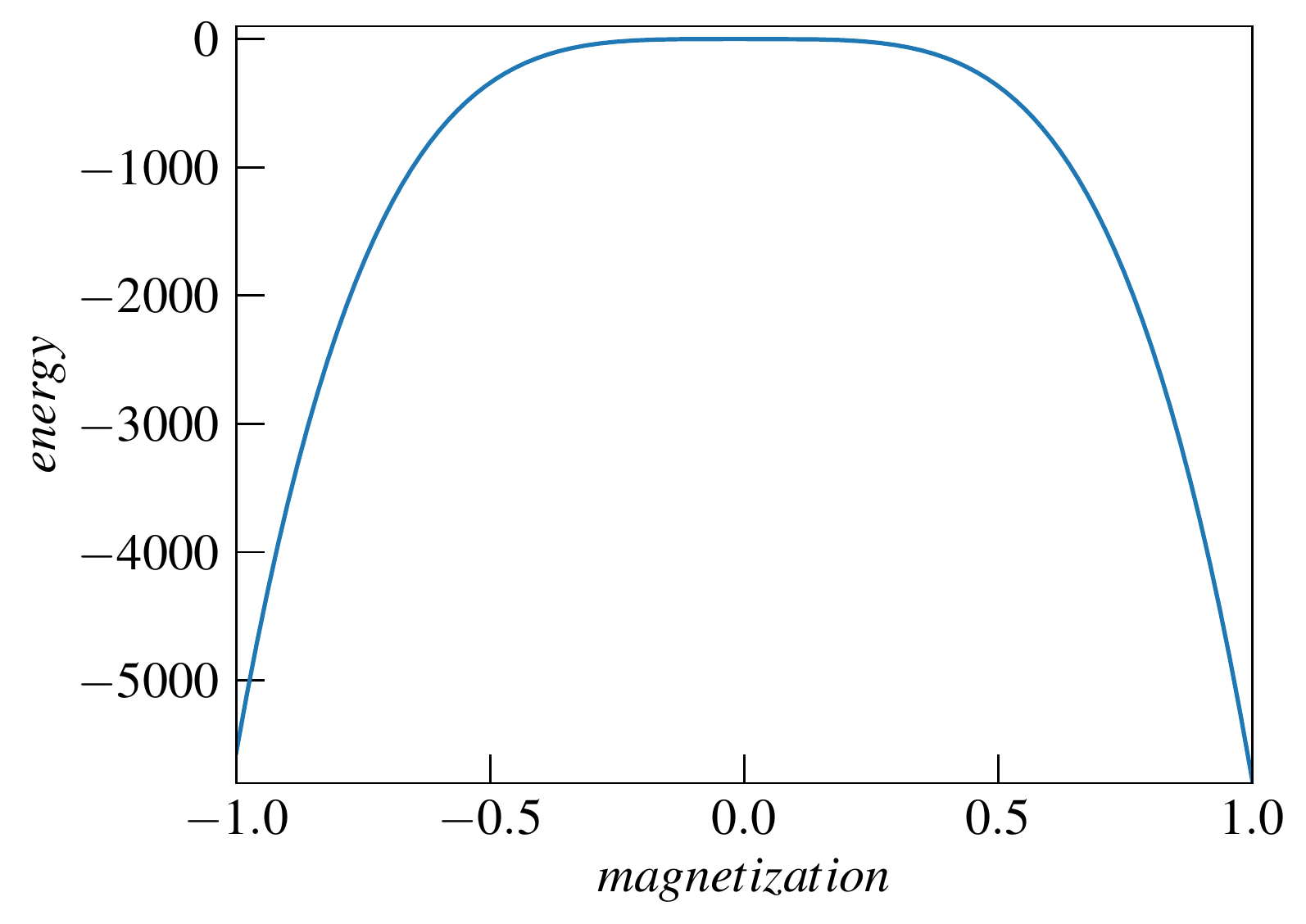}
\caption{\textbf{Energy in thermodynamic limit.} The dependence of the energy on the magnetization for $N_l=109$ logical and thus $N_p=5886$ physical qubits in LHZ is plotted. In the thermodynamic limit $N_p \to \infty$, the behavior of $m^4$ fully dominates the energy term \eqref{eq:energy}.}
\label{fig:figAppendixEnergyTerm}
\end{figure}
With this fact in mind, Fig.~\ref{fig:figAppendixEnergyTerm} shows the energy which approaches a symmetric function due to the fact that the $m^4$ term is dominant for the case of \mbox{$N_p=5886$} physical qubits.\\
For the derivation of the finite-size free-energy expression Eq.~\eqref{eq:eqfreeenergytheory} in LHZ, we need to apply the static approximation $m=m(\alpha)$ for all $\alpha$ and take the reverse operation of the Suzuki-Trotter decomposition for $M \to \infty$ for the expression Eq.~\eqref{eq12:freeenergy}. This gives us the expression,
\begin{widetext}
\begin{align}
f(m,N_p)&=sC\left[\left(3 + \dfrac{6-\sqrt{9+72N_p}}{N_p}\right)m^4+ \left(\dfrac{\sqrt{1+8N_p}-3}{N_p} \right) m^3 \right] \nonumber \\
&-\dfrac{1}{\beta N_p} \sum_{k=1}^{N_p} \ln 2\cosh\beta \sqrt{s^2\left(C\left(4-\dfrac{\sqrt{16+128N_p}+8}{N_p}\right)m^3+C\left(\dfrac{\sqrt{2.25+18N_p}-4.5}{N_p} \right)m^2+J_k\right)^2+h_k^2},
\end{align}
\end{widetext}
where we further use the zero-temperature limit $T \to 0$, i.e., $\beta \to \infty$, and rewrite the sum into an integral for large $N_p$. In this thermodynamic limit $N_p \to \infty$, the actually stepwise function $h_k$ becomes continuous, i.e., $h_k(\tau')$, and, thus, we obtain the free-energy expression for a finite-size system in LHZ as in Eq.~\eqref{HamiltonianLHZInhomogeneous} in the main text.

\begin{figure}
\centering
\includegraphics[width=0.45\textwidth]{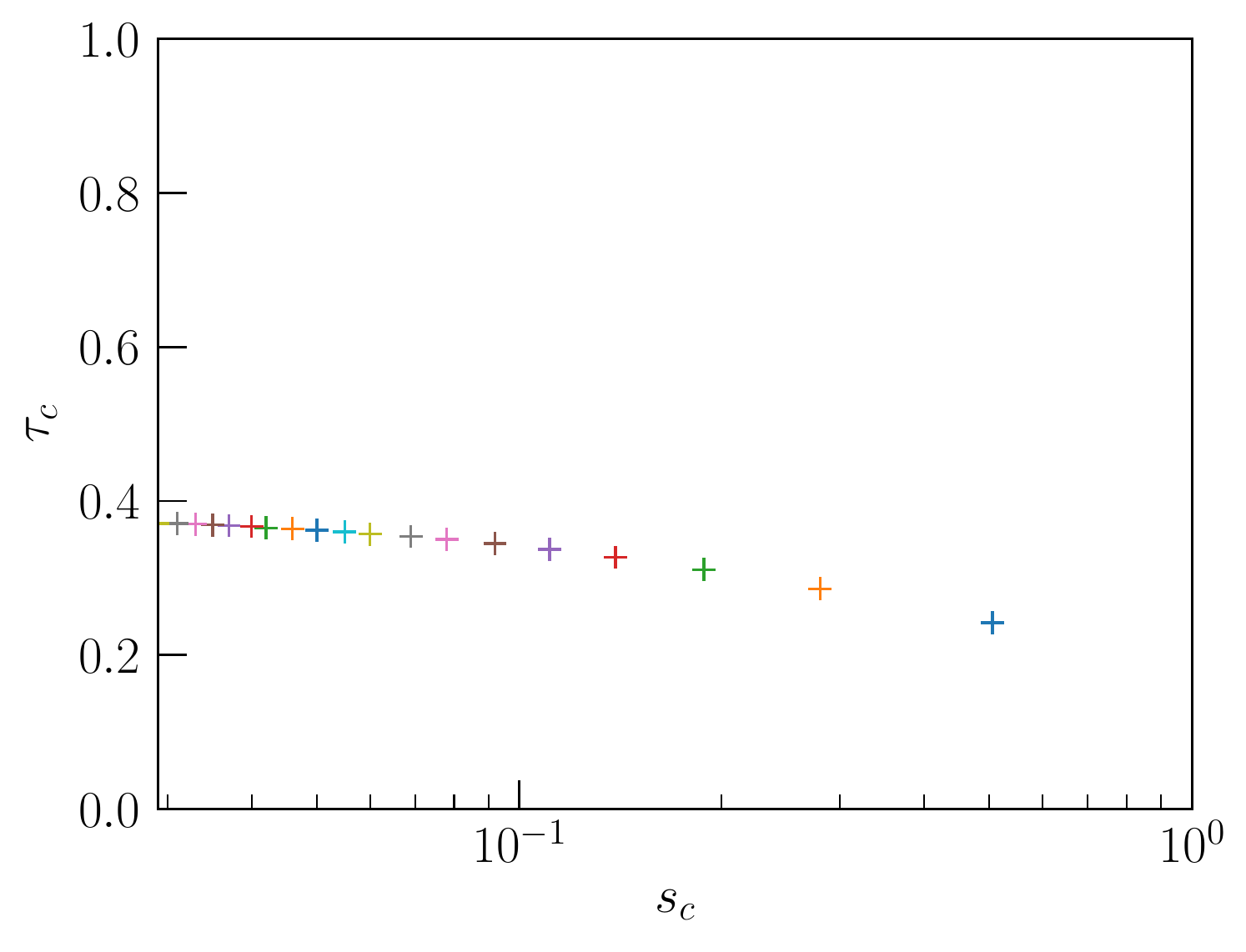}
\caption{\textbf{Scaling with the constraint $C$.} The critical coefficients of finite-size free-energy Eq.~\eqref{eq:Cscaling} in LHZ with randomly chosen instances of interaction strengths $J_k$ for increasing number of $N_l=7$, i.e., $N_p=21$ physical qubits [blue (dark gray) plus on the right] to $N_l=25$, i.e., $N_p=300$ [beige (light gray) plus on the very left] are shown. The scaling of constraint strength for all constraints is chosen to be $C \propto \sqrt{N_p}$.}
\label{fig:AppendixCriticalCoefficients}
\end{figure}

A scaling of the constraint strengths in the form of $C_l \propto \sqrt{N_p}$ may have to be applied in LHZ to suppress any unreasonable solutions in the emerging unreachable sub Hilbert space due to the increase in $N_l$ logical to $N_p \approx N_l^2$ qubits and, thus, increasing Hilbert space.
To account for this, the finite-size free-energy term Eq.~\eqref{eq:eqfreeenergytheory} in LHZ can be rewritten as
\begin{widetext}
\begin{align}
&f(m,s,\tau',J,N_p)=s\left[3m^4 \sqrt{N_p} +\sqrt{8}\left(m^3-3m^4\right)+\dfrac{1}{\sqrt{N_p}} \left(m^4 \left(6-\dfrac{3}{4\sqrt{2}}\right)+m^3\left(\dfrac{1}{4\sqrt{2}}-3\right)\right)-\dfrac{1}{128\sqrt{2}}\dfrac{1}{N_p}\left(m^3-3m^4\right) \right] \nonumber \\
&-\left[\int_0^1 d\tau'\sqrt{s^2\left(4m^3\sqrt{N_p}+\sqrt{8}\left(\dfrac{3}{2}m^2-4m^3\right)-\dfrac{1}{\sqrt{N_p}}\left(8\dfrac{1}{\sqrt{2}}m^3-m^2\left(\dfrac{3}{8\sqrt{2}}-\dfrac{9}{2}\right)\right)-\dfrac{1}{N_p}\left(\dfrac{1.5m^2-4m^3}{128\sqrt{2}}\right)+J\right)^2+h(\tau')^2}\right],
\label{eq:Cscaling}
\end{align}
\end{widetext}
Here, it can be clearly seen that the four-body term dominates with increasing system size. In the second term is a constant offset of the free-energy which stems from the finite-size three-body constraints in LHZ.
The critical coefficients of the finite-size free-energy expression \eqref{eq:Cscaling} in LHZ can, thus, be calculated according to Eq.~\eqref{eq:eqEqSys} in the main text and are plotted in Fig.~\ref{fig:AppendixCriticalCoefficients}.
We can see that the critical coefficients wander from the right of the $s_c-\tau_c$ diagram to the left. The critical coefficients for a system size $N_l=7$ logical spins and, thus, $N_p=21$ physical qubits in LHZ, read $s_c \approx 0.505$ and $\tau_c \approx 0.242$ (blue plus); whereas for a system size $N_l=25$ and, thus, $N_p=300$, the critical coefficients read $s_c \approx 0.029$ and $\tau_c \approx 0.371$ (beige plus). For the thermodynamic limits $N_p \to \infty$ and $C \to \infty$, the critical coefficients will be on the $\tau_c$ axis (i.e., $s_c=0$) so that first-order quantum phase transitions cannot be avoided for any choice of the control parameter $r$.

\subsection*{Additional numerical results}
\subsubsection*{Choice of control parameter $r$}
\begin{figure}
\centering
\includegraphics[width=0.45\textwidth]{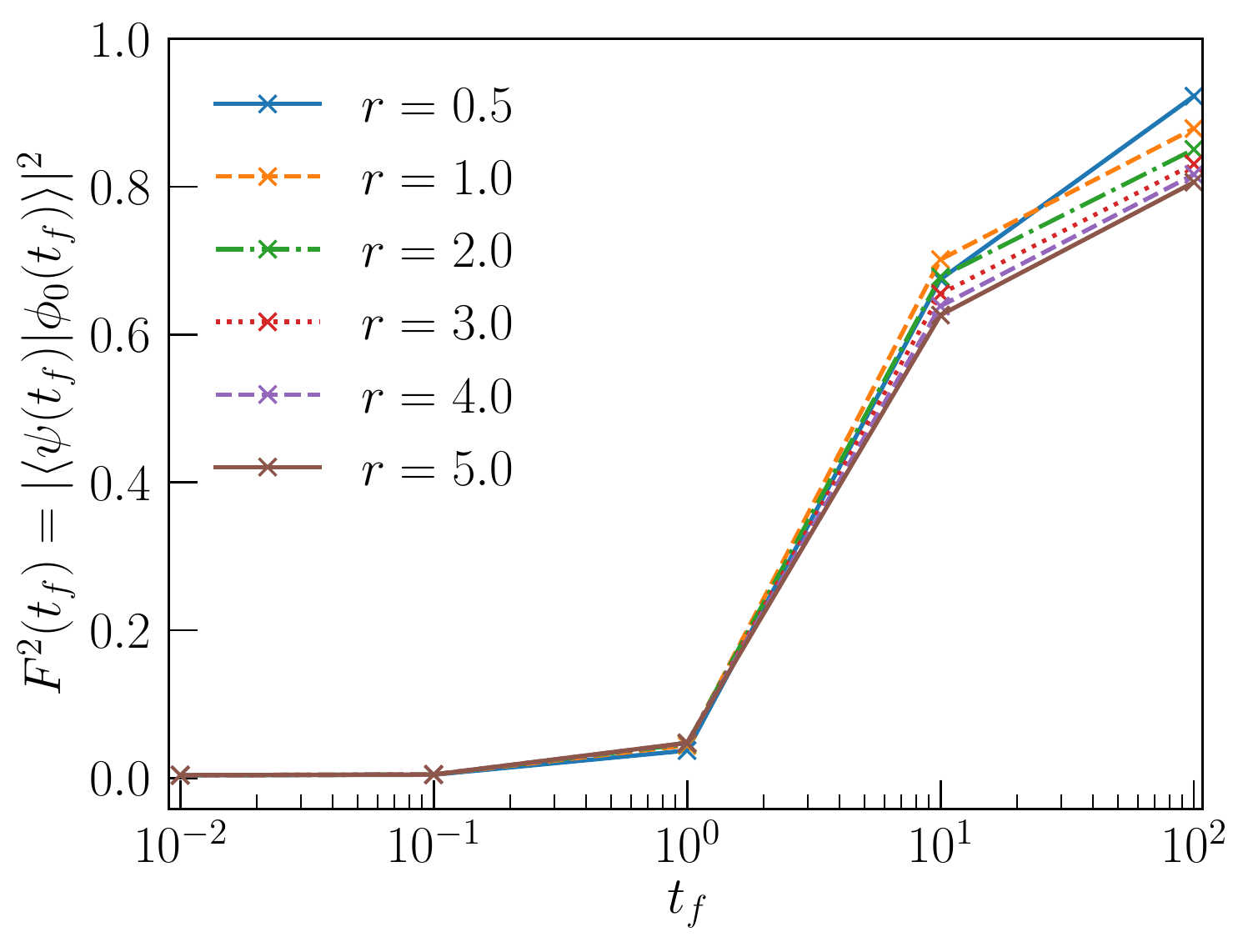}
\caption{\textbf{Control parameter $r$.} The statistics of the final ground-state fidelities $F^2(t_f)=|\langle \psi(t_f)|\phi_0(t_f) \rangle|^2$ of Hamiltonian Eq.~\eqref{HamiltonianLHZInhomogeneous} for an ensemble of 100 instances of randomly uniformly distributed interaction strengths $J_k$ with system size $N_l=4$ logical spins and, thus, $N_p=6$ plus two auxiliary physical qubits and constraint strength $C=2$ for all three constraints and different values of the control parameter $r$ are depicted. The blue (dark gray) upper line corresponds to the value of $r=0.5$ and, subsequently, downwards and the brown lowest solid line corresponds to the value $r=5.0$.}
\label{fig:figComparisonR}
\end{figure}

As mentioned in the main text, the free control parameter $r$ sets the path in the $s-\tau$ diagram one chooses in order to avoid first-order quantum phase transitions. Its choice can also further enhance the performance of quantum annealing by not only avoiding first-order QPTs, but also enlarging the probability of finding the ground-state of our problem Hamiltonian to be solved.\\
Figure \ref{fig:figComparisonR} depicts the statistics of the squared final ground-state fidelity $F^2(t_f)=|\langle \psi(t_f)|\phi_0(t_f) \rangle|^2$ for Hamiltonian Eq.~\eqref{HamiltonianLHZInhomogeneous} with system size $N_p=6$ plus two auxiliary physical qubits in LHZ for an ensemble of 100 randomly chosen instances of $J_k$ and different values of the control parameter $r$.
As can be seen here, an appropriate choice of the control parameter $r$ can enlarge the squared final ground-state fidelity of the inhomogeneously driven Hamiltonian considerably. For a value of $r=0.5$, the final ground-state fidelity squared reaches its maximum.

\begin{figure}[htbp]
\centering
\includegraphics[width=0.45\textwidth]{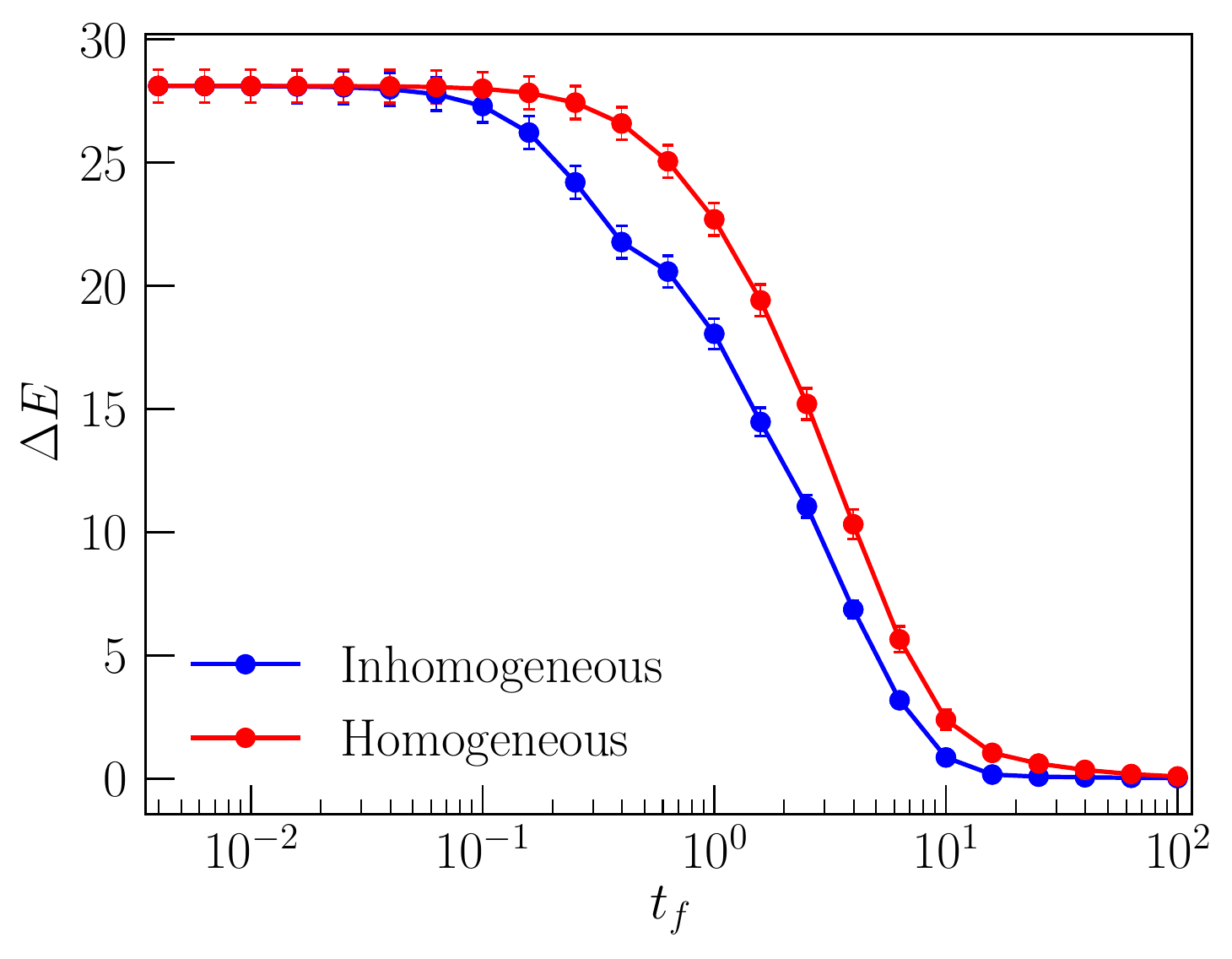}
\caption{\textbf{Excess energies for sweeps with homogeneous and inhomogeneous driving.} The statistics of the excess energies for an ensemble of 100 randomly uniformly distributed interaction strengths $J_k$ with system size $N_l=4$ logical qubits and thus $N_p=6$ physical qubits plus two auxiliary qubits and constraint strength $C_l=2$ for all three constraints for homogeneous \eqref{eq:eqHamiltonianLHZ} and inhomogeneous Hamiltonian Eq.~\eqref{HamiltonianLHZInhomogeneous} with parameter $r=0.5$ is depicted.}
\label{fig:figExcessEnergies}
\end{figure}

The excess energy is another measure for the performance of quantum annealing as it gives rise to the amount of transitions to higher excited eigenstates that have occurred during quantum annealing sweeps.\\
Figure \ref{fig:figExcessEnergies} depicts the excess energies of the homogeneously driven Hamiltonian \eqref{eq:eqHamiltonianLHZ} and inhomogeneously driven Hamiltonian Eq.~\eqref{HamiltonianLHZInhomogeneous} with parameter $r=0.5$ and where the parameters of LHZ are as described in the main text.
The excess energies of the inhomogeneously driven Hamiltonian \eqref{HamiltonianLHZInhomogeneous} are smaller than for the homogeneously driven Hamiltonian \eqref{eq:eqHamiltonianLHZ} in LHZ, meaning that less transitions to higher excited states have occurred during these sweeps.

\subsubsection*{Different orders of inhomogeneous driving}
\begin{figure}
\centering
\includegraphics[width=\textwidth]{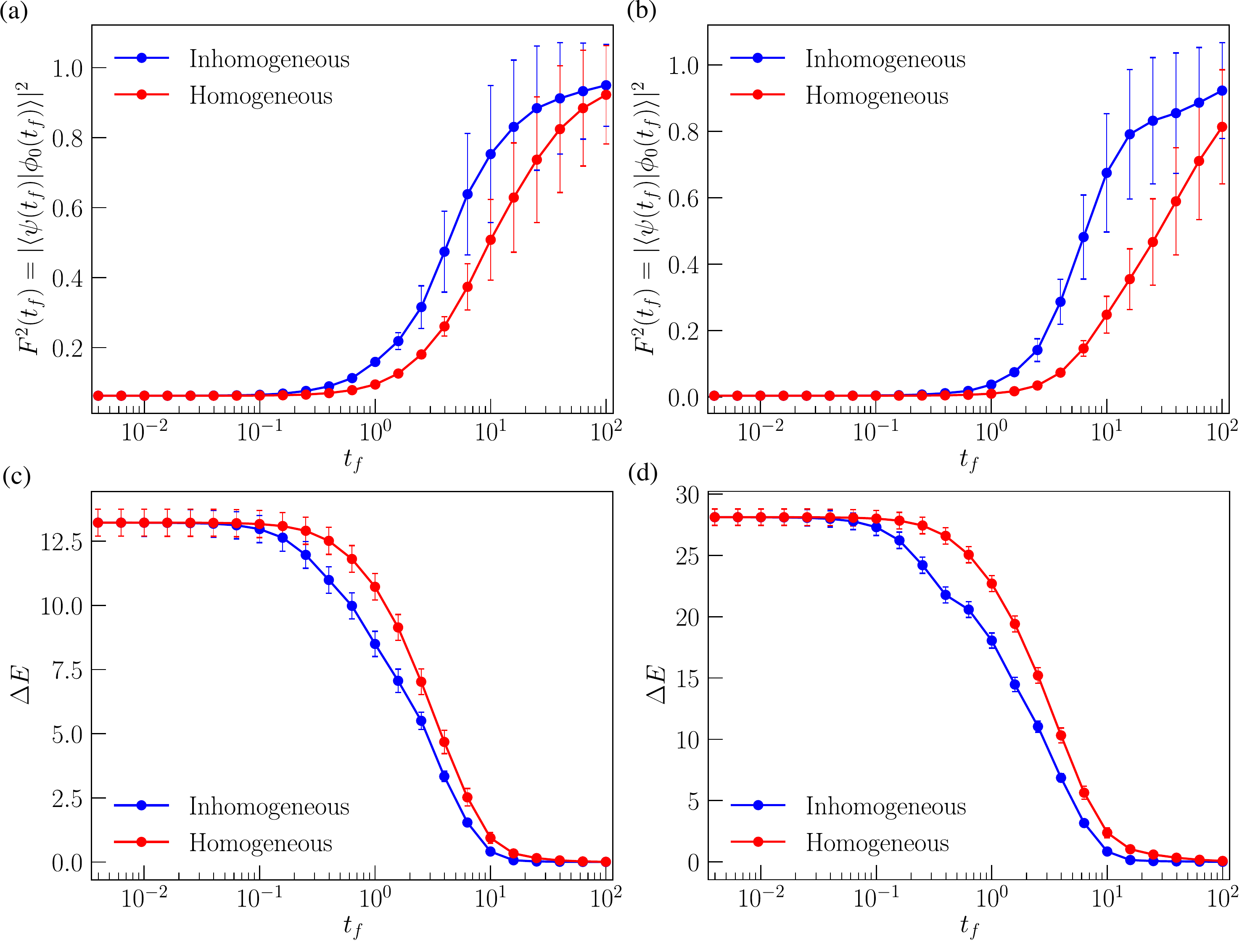}
\caption{\textbf{Statistics of descending order.} (a) and (c) show the statistics of the final ground-state fidelity of the homogeneous \eqref{eq:eqHamiltonianLHZ} and inhomogenous Hamiltonian Eq.~\eqref{HamiltonianLHZInhomogeneous} with parameter $r=0.5$ for system sizes $N_p=3$ and $N_p=6$ physical qubits, respectively, each for an ensemble of 100 uniformly distributed instances of interaction strength $J_k$. (b) and (d) show the statistics of the corresponding excess energies for $N_p=3$ and $N_p=6$ physical qubits, respectively, and same ensemble of $J_k$ and control parameter $r=0.5$ as for (a) and (c).}
\label{fig:figStatisticsDescending}
\end{figure}

An interesting question arises whether the order in which we switch off the transverse field of each qubit is of matter for our theory.
Figure \ref{fig:figStatisticsDescending} depicts the statistics of final ground-state fidelity and excess energies of the descending order of inhomogeneously driven transverse fields of the qubits over an ensemble of 100 uniformly distributed interaction strengths $J_k$ for system sizes $N_p=3$ plus one auxiliary and $N_p=6$ plus two auxiliary physical qubits, respectively. Here, we first switch off the transverse field of the qubit at the top of the triangular LHZ structure and the qubits 1 in the left lower row and auxiliary qubits in the last row.
We see the same results as in the case of ascending order from the main text with function $h_k(s,r)$ as in Figure \ref{fig:figAppendix1}, meaning that the order in which we switch off the transverse field of the qubits does not matter for the efficiency of our method.

\subsubsection*{Energy spectrum}
\begin{figure}
\centering
\includegraphics[width=\textwidth]{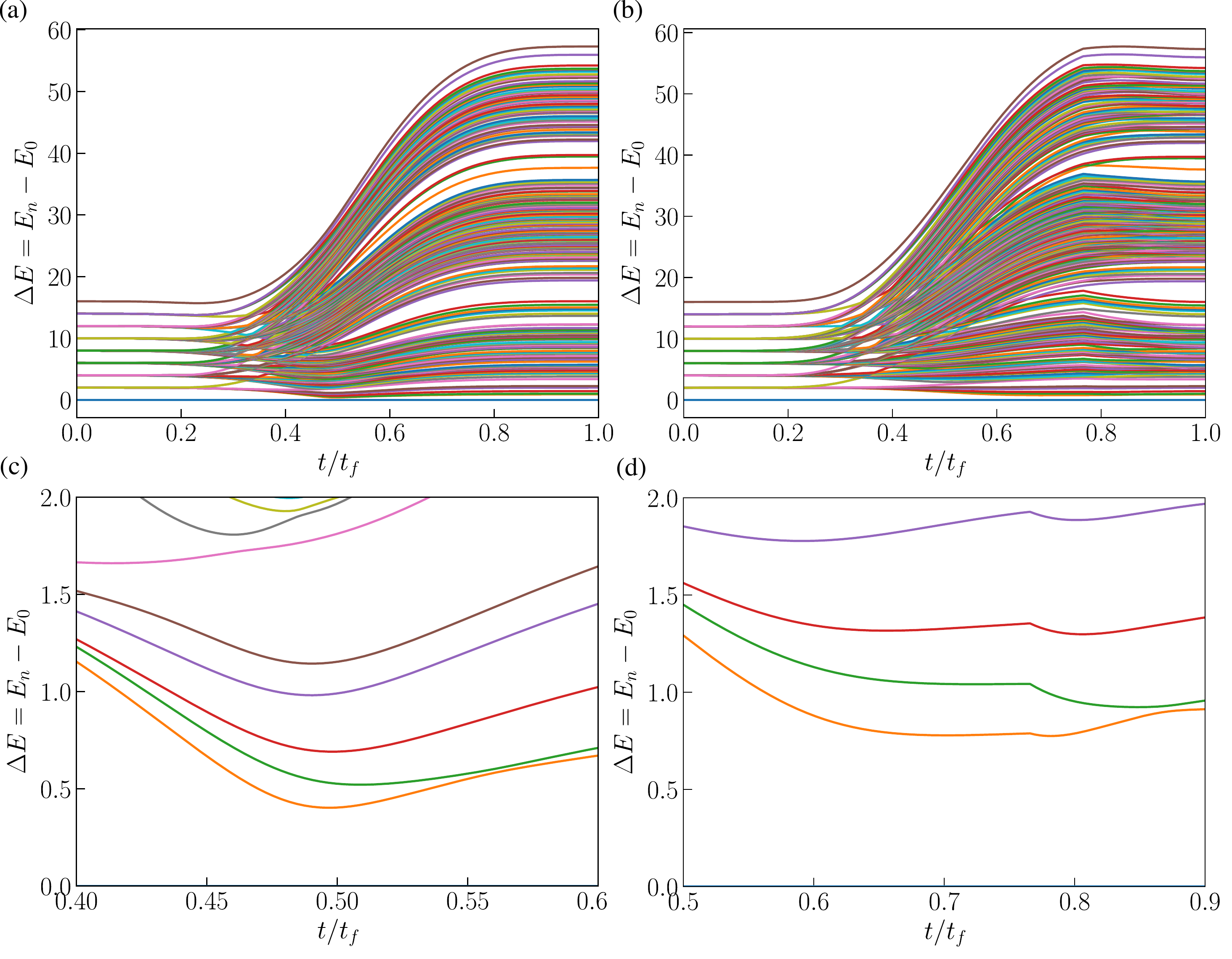}
\caption{\textbf{Energy Spectrum.} (a) and (c) show the energy spectrum of the homogeneous Hamiltonian Eq.~\eqref{eq:eqHamiltonianLHZ} with $N_p=6$ physical qubits for a randomly chosen instance of interaction strength $J_k$. (b) and (d) show the energy spectrum of the inhomogeneous Hamiltonian Eq.~\eqref{HamiltonianLHZInhomogeneous} with parameter $r=0.5$ with $N_p=6$ physical qubits with the same instance $J_k$. The constraint strength is $C_l=2$ for all three constraints.}
\label{fig:EnergySpectrum}
\end{figure}

As we can enlarge the minimal energy gap by inhomogeneous driving of the transverse field in LHZ, we are interested in the energy spectrum during the whole sweep.
Figure \ref{fig:EnergySpectrum} shows the energy spectra of the homogeneous \eqref{eq:eqHamiltonianLHZ} and inhomogenous Hamiltonian Eq.~\eqref{HamiltonianLHZInhomogeneous} with parameter $r=0.5$ for system size $N_p=6$ plus two auxiliary physical qubits and a randomly chosen instance $J_k$ with constraint strength $C_l=2$ for all constraints. We can see that the minimal energy gap is enlarged and shifted in time from around $t/t_f \approx 0.5$ for the homogeneous Hamiltonian to $t/t_f \approx 0.8$ for the inhomogeneous Hamiltonian.

\onecolumngrid

\twocolumngrid

\end{document}